\documentclass{PoS}
\usepackage{lineno}
\usepackage{float}
\title{Searches for strong production of supersymmetry in CMS}

\ShortTitle{Searches for strong production of supersymmetry in CMS}

\author{\speaker{Tamas Almos Vami} for the CMS Collaboration\thanks{On behalf of the CMS SUSY razor boost analysis team. Special thanks to Janos Karancsi and Sezen Sekmen.}\\
        Wigner Research Centre for Physics\\
        E-mail: \email{tamas.almos.vami@cern.ch}}

\abstract{Searches for production of supersymmetric partners of gluons and quarks with the CMS experiment at CERN's LHC have excluded these particles for masses up to $\approx2$ TeV. The talk will present results and show the analysis techniques in these searches, with an emphasis on the use of razor variables for the discrimination between standard model backgrounds and signal.}

\FullConference{Corfu Summer Institute 2018 "School and Workshops on Elementary Particle Physics and Gravity"\\
		(CORFU2018)\\
		31 August - 28 September, 2018\\
		Corfu, Greece}

\begin{document}

\section{Introduction}
It is known that the Standard Model of particle physics (SM) cannot be the final theory of the Universe as it does not explain the matter/antimatter asymmetry, it does not provide any candidate on Dark Matter, it does not give any reason why the mass of the Higgs boson is so low with respect to the Planck mass and it does not unify the gauge couplings. Extending the SM with new particles (superpartners) based on a new symmetry called supersymmetry (SUSY) \cite{ZuminoSUSY} seems to solve these problems theoretically \cite{GUT,WittenSUSYBreaking,Predictions,EllisBigBang,SUSYDM}. Superpartners of the SM fermions are bosons while the SM bosons have fermionic partners.

\begin{figure}[H]
    \centering
    \includegraphics[width=\textwidth]{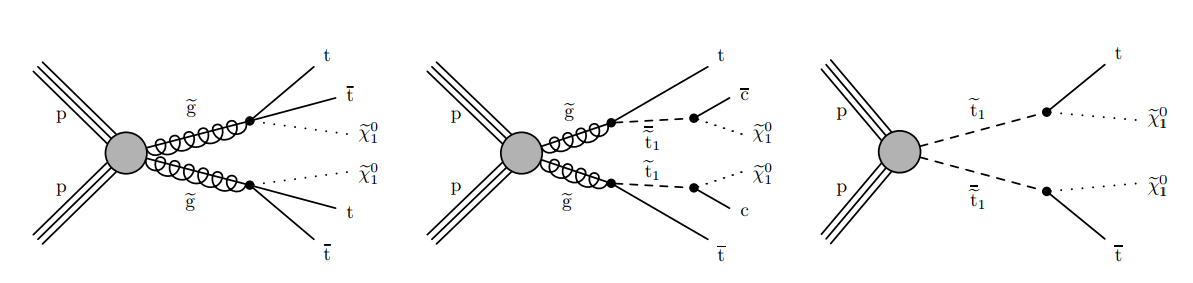}
    \caption{Diagrams for the SMSs considered in this analysis: (left) Gluino pair production decaying to two top quarks and the LSP, (middle) Gluino pair production decaying to a top quark and a low mass top squark, which subsequently decays to a charm quark and the LSP. (right) Top squark pair production decaying to a top quark and the LSP.}
    \label{fig:SMSs}
\end{figure}

In this paper, searches for strong production SUSY using razor variables \cite{razorArticle} with data from the CMS detector in 2016 are presented for an integrated luminosity of 35.9 $\mathrm{fb}^{-1}$. We divide the searches into gluino searches, that are the superpartners of gluons and into squark searches that are the superpartners of quarks. 
We have analyzed the following three decay channels in the framework of simplified models (SMS) \cite{SMS}, which are shown in Figure \ref{fig:SMSs}.
\begin{itemize}
    \item Gluino pair production with gluinos decaying to two top quarks and Lightest SUSY Particle (LSP)
    \item Gluino pair production with gluinos decaying to a top quark and a low mass top squark, which subsequently decays to a charm quark and the LSP 
    \item  Top squark pair production with top squarks decaying to a top quark and the LSP.
\end{itemize}

This is the first search for SUSY from the CMS experiment that simultaneously studied and combined both Lorentz-boosted and "nonboosted" (resolved) event categories.
A more detailed version of these results can be found in \cite{SUS-16-017}.

\section{The CMS detector}
The central feature of the CMS apparatus is a superconducting solenoid of 6 m internal diameter, providing a magnetic field of 3.8 T. Within the solenoid volume are a silicon pixel and strip tracker, a lead tungstate crystal electromagnetic calorimeter (ECAL), and a brass and scintillator hadron calorimeter (HCAL), each composed of a barrel and two endcap sections. Forward calorimeters extend the pseudorapidity coverage provided by the barrel and endcap detectors. Muons are detected in gas-ionization chambers embedded in the steel flux-return yoke outside the solenoid. A more detailed description of the CMS detector, together with a definition of the coordinate system used and the relevant kinematic variables, can be found in Ref.~\cite{Chatrchyan:2008zzk}.

Events of interest are selected using a two-tiered trigger system~\cite{Khachatryan:2016bia}. The first level (L1), composed of custom hardware processors, uses information from the calorimeters and muon detectors to select events at a rate of around 100 {kHz} within a time interval of less than 4$\mu$s. The second level, known as the high-level trigger (HLT), consists of a farm of processors running a version of the full event reconstruction software optimized for fast processing, and reduces the event rate to around 1 {kHz} before data storage.

\section{Object reconstruction}

The particle-flow (PF) algorithm~\cite{CMS-PRF-14-001} aims to reconstruct and identify each individual particle in an event, with an optimized combination of information from the various elements of the CMS detector. The energy of photons is obtained from the ECAL measurement. The energy of electrons is determined from a combination of the electron momentum at the primary interaction vertex as determined by the tracker, the energy of the corresponding ECAL cluster, and the energy sum of all bremsstrahlung photons spatially compatible with originating from the electron track. The energy of muons is obtained from the curvature of the corresponding track. The energy of charged hadrons is determined from a combination of their momentum measured in the tracker and the matching ECAL and HCAL energy deposits, corrected for zero-suppression effects and for the response function of the calorimeters to hadronic showers. Finally, the energy of neutral hadrons is obtained from the corresponding corrected ECAL and HCAL energies.

Jets are clustered from PF candidates using the anti-kT algorithm with a distance parameter of 0.4. All jets in this analysis are required to have $p_T > 30$ GeV and pseudorapidity $|\eta | < 2.4$. Tagging b-jets is crucial for this analysis, too. The "medium" and the "loose" working points of the combined secondary vertex (CSVv2) b jet tagger were used, which essentially uses an inclusive vertex finder to select b jets \cite{bjets}.

In order to identify the Lorentz-boosted W bosons (top quarks) the anti-kT algorithm with a distance parameter of 0.8 was used with $p_T> 200 (400)$ GeV. Identification of boosted jets is done using jet mass, the N-subjettiness
variables~\cite{Thaler:2010tr}, and subjet b tagging for top quarks. N-subjettiness is defined as
\begin{equation}
    \tau_N = \frac{1}{R_0\sum_k p_{T,k}} \sum_k p_{T,k} \,min (\Delta R_{1,k},\Delta R_{2,k},\dots,\Delta R_{N,k})
\end{equation}
where N denotes candidate axes for subjets, k runs over all constituent particles, and $R_0$ is the clustering parameter of the original jet, and $\Delta R_{N,k}$ is the distance from constituent particle k to subjet n. The N-subjettiness variable is used to evaluate the consistency of a jet with having N subjets.

Hadronic decays of top quarks are identified using the ratio between 3-subjettiness and 2-subjettiness, $\tau_{32}=\tau_{3}/\tau_{2}$, and the groomed jet mass ~\cite{Butterworth:2008iy,Larkoski:2014wba}. 
Hadronic decays of $W$ bosons are identified using the ratio between 2-subjettiness and 1-subjettiness, $\tau_{21}=\tau_{2}/\tau_{1}$, and the groomed jet mass. 

The missing transverse momentum vector $\vec{p}_T^{miss}$
is defined as the projection of the negative
vector sum of the momenta of all reconstructed PF candidates on the plane perpendicular to
the beams. Its magnitude is referred to as ${p}_T^{miss}$.

\section{Event categorization, backgrounds and systematic uncertainties}
Events are categorized according by the number of leptons, jets, and b-tagged jets identified. 
There is a category for events that contain one and only one charged lepton which is called the "one-lepton" category. The zero-lepton category has several subcategories, for example if the event contains boosted hadronic W boson or top quark decays then it will belong to the "boosted" event category, otherwise it will go to the "multijet" category. These subcategories are further diveded to smaller cases based on the number of jets in the event.

As a next step we group the physics objects into two distinct hemispheres called megajets, whose four-momenta are defined as the vector sum of the four-momenta of the physics objects in each hemisphere. 

The SUSY search is performed in bins of the razor variables $M_R$ and $R^2$, defined as

\begin{equation}
    M_R = \sqrt{(|\vec{p}^{j_1}|+|\vec{p}^{j_2}|)^2 - (p_z^{j_1}+p_z^{j_2})^2 }
\end{equation}
and
\begin{equation}
    M_T^R = \sqrt{\frac{p_T^{miss}(p_T^{j_1}+p_T^{j_1})-\vec{p}_T^{miss}(\vec{p}_T^{j_1}+\vec{p}_T^{j_1})}{2}}
\end{equation}

where $j_1$ and$j_2$ refers to the megajets.

From these we define R
\begin{equation}
    R= \frac{M_T^R }{M_R }
\end{equation}

Due to the R-parity conservation we expect the SUSY particles to be produced in pairs. The $M_R$ variable gives the mass splitting between the pair-produced particle and the LSP and it has a peaking structure, while for background it is an exponentially decaying function. The variable R shows the imbalance between the visible and invisible decay products. The combination of the two variables
provide powerful discrimination between the SUSY signal and SM backgrounds.

The main background is coming from W($\ell \nu$)+jets, Z($\nu\bar{\nu}$)+jets, $t\bar{t}$ and QCD multijet production. In case of zero b-tagged jets W($\ell \nu$)+jets,  Z($\nu\bar{\nu}$)+jets  while for more b-tagged jets the $t\bar{t}$ is dominant. 

The SUSY signal was produced at leading order with MADGRAPH5\_aMC@NLO 2.2.2 interfaced with PYTHIA V8.205 for fragmentation and parton showering, and matched to the matrix element kinematic configuration using the MLM algorithm. The background processes were generated at next-to-leading order (NLO) with MADGRAPH5\_ aMC@NLO 2.2.2 (W+jets, s-channel single top quark, ttW, ttZ processes) or with POWHEG v2.0 (tt+jets, t-channel single top quark, and tW production), both interfaced with PYTHIA V8.205.
All simulated events include the effects of multiple pp collisions within the event (also called as pileup).

The background prediction strategy uses control regions to isolate each background process, address any deficiencies of the MC simulation in a data driven way, and estimate systematic uncertainties in the expected event yields. The QCD multijet, $t\bar{t}$-jet, W+jets,Z+jets are estimated from the control regions, while the rest uses the MC simulations.

Systematic uncertainties can come from the limited accuracy of measurement, from the data-driven background prediction methodology and from the fast simulated prediction of the signal. The b mistagging efficiency has the biggest impact (from 2-20\%) which effects both signal and the background estimate. Higher-order corrections can lead to a 10-20 \% uncertainty, while the fast simulation corrections have an impact of 1-5\%.

\section{Results and interpretation}
The observed data yields in the signal regions are statistically consistent with the background prediction from SM processes.

\begin{figure}[H]
    \centering
    \includegraphics[width=\textwidth]{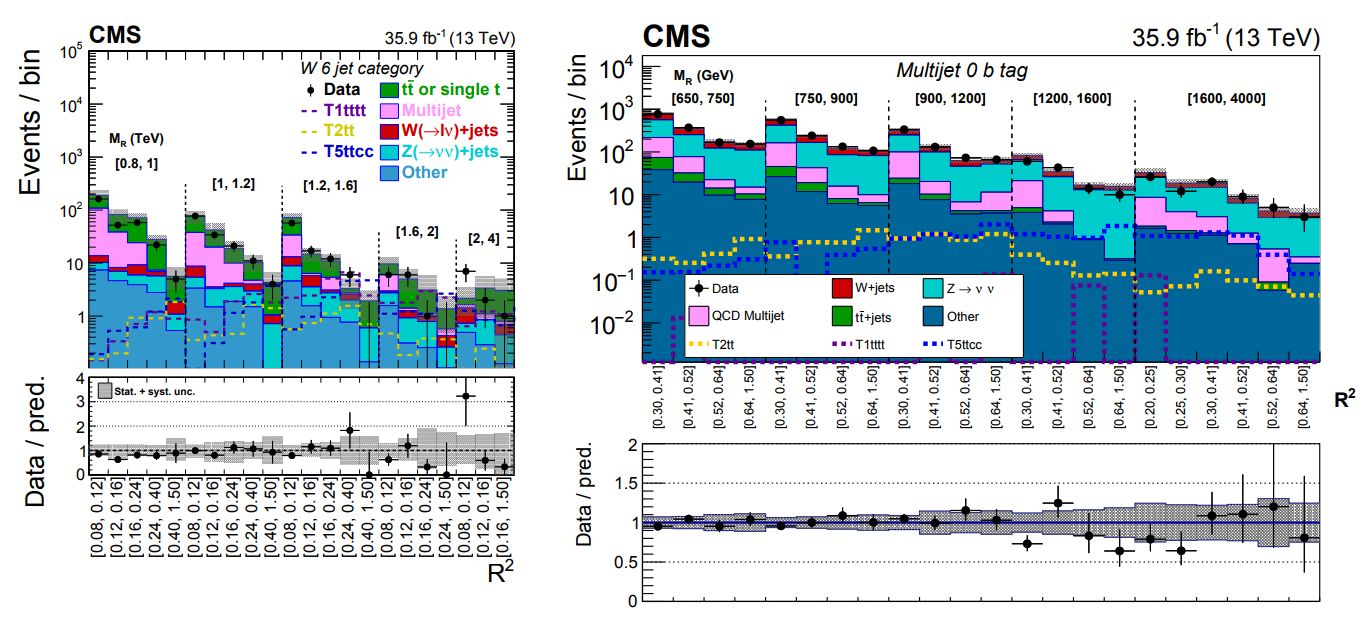}
    \caption{Left: The $M_R-R^2$ distribution in the zero lepton category with a boosted W with 6 jets. Right: The $M_R-R^2$ distribution in the nonboosted zero lepton category with multijets but no b-tagged jet. \cite{SUS-16-017}}
    \label{fig:BkgVsMC}
\end{figure}

One $M_R-R^2$ distribution in the zero-lepton boosted category (in this case the W category with 6 jets) is shown on the left part of Figure \ref{fig:BkgVsMC},
while a result for the zero-lepton nonboosted category (specifically the multijet with 0 b-tagged jets) is shown in the right part of Figure \ref{fig:BkgVsMC}. The two-dimensional $M_R-R^2$ distribution is shown in a one-dimensional representation, with each $M_R$ bin denoted by the dashed lines and labeled, and each $R^2$ bin labeled in the bottom. The background labeled as "Other" includes single top quark production, diboson production, associated production of a top quark pair and a W or Z boson, and triboson production.

The ratio of data to the MC simulation prediction is shown on the bottom box, with the statistical uncertainty expressed using error bars around the data point and the systematic uncertainty of the background prediction represented by the shaded region. 
Similar plots exist for the other categories, too.

Following the LHC CLs procedure \cite{CLs}  we set upper limits on the production cross sections of various SUSY simplified models. We used the profile likelihood ratio test statistic and the asymptotic formula to evaluate the 95\% confidence level (CL) observed and expected limits on the production cross section.

\begin{figure}[h!]
    \centering
    \includegraphics[width=\textwidth]{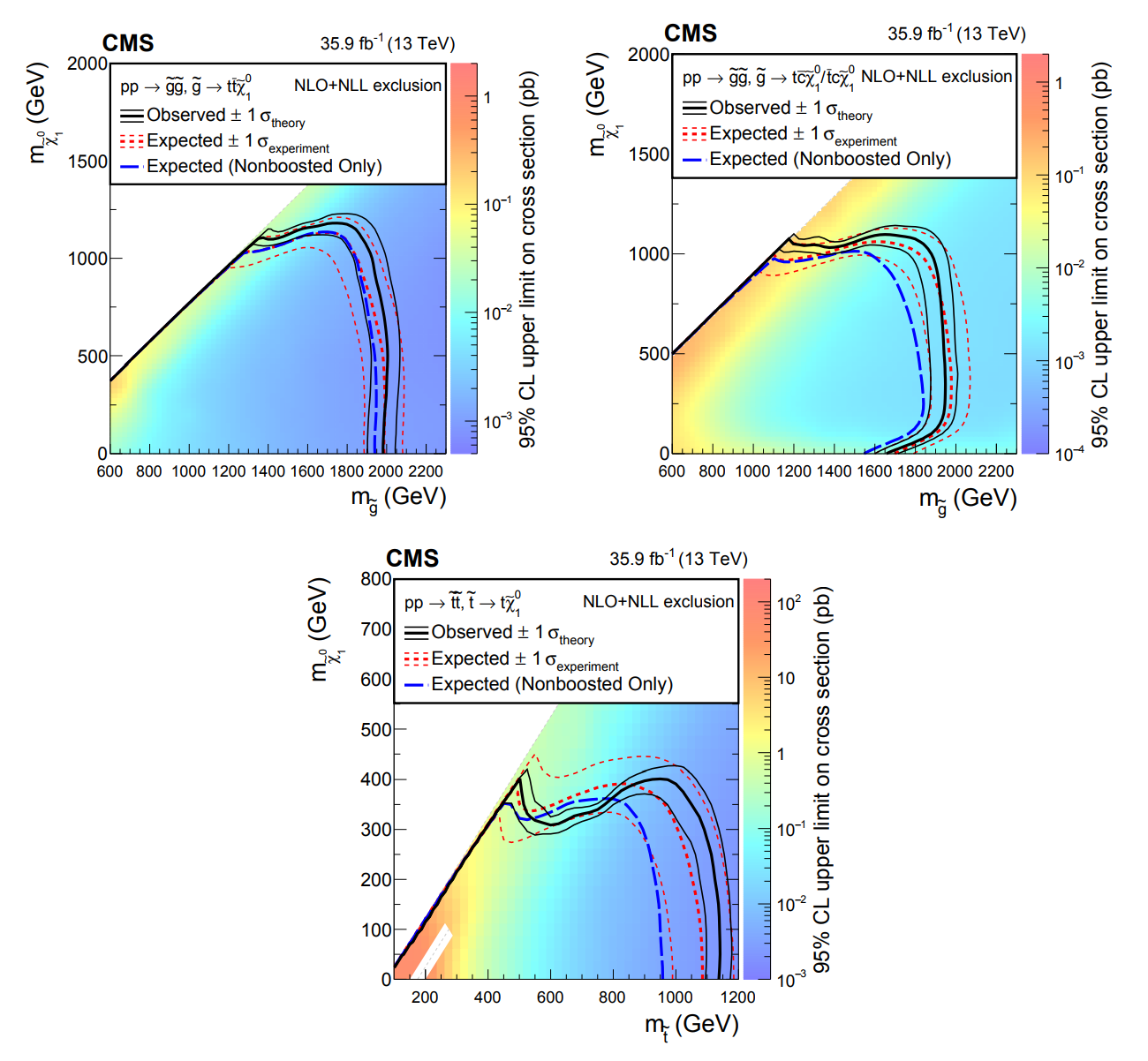}
    \caption{Left: Expected and observed 95\% CL limits on the production cross section for pairproduced
gluinos each decaying to the LSP and top quarks. Right: Expected and observed 95\% CL limits on the production cross section for pairproduced
gluinos each decaying to a top quark and a low mass top squark that subsequently decays to a charm quark and the LSP. The bottom plot shows the expected and observed 95\% CL limits on the production cross section for pairproduced squarks each decaying to a top quark and the LSP. \cite{SUS-16-017}}
    \label{fig:Results1}
\end{figure}

In the scenario of pair produced gluinos decaying to two top quarks and the LSP, Figure \ref{fig:Results1} shows the expected and observed limits as a function of gluino and LSP masses. In this simplified model, we exclude gluino masses up to 2.0 TeV for LSP mass below 700 GeV.
The limit for gluinos decaying to a top quark and a low mass top squark that subsequently decays to a charm quark and the LSP, is shown in the right plot of Figure \ref{fig:Results1}. For this simplified model, we exclude gluino masses up to 1.9 TeV for LSP mass above 150 and below 950 GeV, extending the previous best limits \cite{PrevResults1} from the CMS experiment by about 100 GeV in the gluino mass. The white diagonal band corresponds to the region $|m_{\tilde{t}}-m_t-m_{\tilde{\chi}_1^0}| < 25\, GeV$. In this region the signal acceptance depends strongly on the $\tilde{\chi}_1^0$ mass and thus it is difficult to model.

Finally, we consider pair produced top squarks decaying to the top quark and the LSP. The expected and observed limits are shown in Figure \ref{fig:Results1}, and we exclude top squark masses up to 1.14 TeV for LSP mass below 200 GeV, extending the previous best limits \cite{PrevResults2} from the CMS experiment by about 20 GeV.
In each exclusion limit plot, the dashed blue line represents the expected limit obtained using data from the nonboosted categories only. 
We can observe that adding the boosted categories clearly improves the sensitivity for the signal models presented here.

\section{Conclusions}

An inclusive search for supersymmetry in one lepton and fully hadronic final states using the razor variables and boosted object tagging techniques was presented based on CMS data from 2016 with 35.9 $\mathrm{fb}^{-1}$ of proton-proton collisions at $\sqrt{s}=13\, \mathrm{TeV}$. The combination of both boosted and resolved event categories improved the exclusion limits for gluino and top squark pair production scenarios considered here. Data are observed to be consistent with the SM expectation. Limits on the gluino mass extend to 2.0 TeV while limits on top squark masses reach 1.14 TeV.

\end{document}